\documentclass[prb,amsfonts,amssymb,floats,twocolumn,aps,floatfix,nobalancelastpage]{revtex4-2}
\usepackage{amsmath}
\usepackage{mathtools}
\usepackage{float}
\usepackage[caption = false]{subfig}
\usepackage[dvipsnames]{xcolor}
\usepackage{braket}
\usepackage{bm}
\usepackage{dsfont}
\usepackage[export]{adjustbox}
\usepackage{graphicx}
\usepackage{tikz}
\usepackage{placeins}
\usepackage{bbm}
\usepackage{lipsum}
\usepackage{xfrac}
\usepackage[version=4]{mhchem}
\allowdisplaybreaks
\graphicspath{{./figs}}

\begin{document}

\title{Series expansion studies of the $J_1$-$J_2$-Heisenberg bilayer}
\author{Erik Wagner}
\affiliation{Institute for Theoretical Physics,
Technical University Braunschweig, D-38106 Braunschweig, Germany}
\author{Wolfram Brenig}
\affiliation{Institute for Theoretical Physics,
Technical University Braunschweig, D-38106 Braunschweig, Germany}

\date{\today}

\begin{abstract}
We study a bilayer of the frustrated $J_1$-$J_2$ Heisenberg-model on the square
lattice. Starting from the dimer limit at strong interlayer coupling, we perform
series expansions using the perturbative Continuous Unitary Transformation,
based on the flow equation method, in order to determine the spectrum up to the
two-triplon sector.  From the one-triplon dispersion we obtain quantum critical
lines for transitions from the dimer phase into either N\'eel or collinear
magnetic order. For low to intermediate frustration these transitions are
consistent with existing findings, based on the magnetic phases. In the region
of strongest frustration, i.e. $J_2/J_1\sim 0.5$, we provide an estimate for the
stability of the anticipated single-layer quantum spin-liquids against finite
interlayer coupling.  In the two-triplon sector we find a set of well defined
(anti)bound states, which can be classified according to total spin and in-plane
rotational symmetry. For vanishing frustration these states agree with previous 
series expansion analysis.  For $J_2/J_1\gtrsim 0.5$ we provide evidence for a 
close-by condensation of one-triplon and two-triplon singlet bound states, 
suggesting that between the dimer and the collinear state, additional phases may 
intervene.
\end{abstract}

\maketitle

\section{Introduction}
\label{sec:intro}

Phases and excitations of quantum magnets are among the keys to understand
correlated electron systems \cite{Sachdev2008}. Stepping beyond conventional
long-range magnetic order, exchange frustration is a prime ingredient to
achieve novel states of matter in these magnets, displaying, e.g., spin liquid
behavior, topological order, and exotic excitations \citep{Kitaev2006,
Balents2010, Henley2010, Castelnovo2012, Misguich2012, Savary2016,
Sachdev2018}. In this context, the planar antiferromagnetic $J_1$-$J_2$
Heisenberg-model on the square lattice (J1J2HM) \cite{Chandra1988} is one of
the pillars of frustrated quantum magnetism.  While first analysis of this
model dates back several decades, basic properties, like parts of the quantum
phase diagram still remain open issues. Classically, the ground state is a
N\'eel state for $\kappa = J_2/J_1 < 1/2$, and comprises two inter-penetrating
N\'eel states with $\sqrt{2}\times\sqrt{2}$ structure for $\kappa = J_2/J_1 >
1/2$. The relative degeneracy of the latter N\'eel vectors is lifted by
"order-by-disorder" \cite{Henley1989, Moreo1990, Chandra1990}, leading to
columnar order with an Ising, i.e. $Z_2$, symmetric order parameter of the
pitch vector at $(0,\pi)$ and $(\pi,0)$. At $\kappa = 1/2$ the Luttinger-Tzia
method \cite{Luttinger1946} results in macroscopic degeneracy of the classical
ground state due to line minima of the energy versus the ordering pitch
vector. Moreover, the leading order zero-temperature $1/S$ corrections to the
order parameter diverge at $\kappa = 1/2$ \cite{Chandra1988}.

Turning to the quantum limit, i.e., $S=1/2$, solid evidence has been gathered,
that for $0.5\lesssim k \lesssim 0.6$ no magnetic order exists. This has been
collected from a plethora of approaches, some of which include exact
diagonalizations \cite{Elbio1989, Didier1991, Didier1996, Didier2006}, series
expansions \cite{Oitmaa1996, Rajiv1999, Sirker2006}, coupled-cluster theory
\cite{Darradi2008, Richter2015, Papastathopoulos2022}, variational methods
\cite{Mezzacapo2012, Ren2014}, density-matrix renormalization group (DMRG)
\cite{Jiang2012, Gong2014, Wang2018}, (infinite) projected entangled-pair state
((i)PEPS) \cite{Haghshenas2018, Hasik2021, Liu2022}, functional-renormalization
group (fRG) \cite{Hering2019, Roscher2019}, perturbative analysis
\cite{Zhitomirsky1996, Doretto2014}, and variational Monte Carlo (VMC)
calculations \cite{Hu2013, Morita2015, Ferrari2020, Nomura2021}. Instead of
magnetic order, these approaches have predicted various novel ground states,
including plaquette valence-bond crystals (PVBC), columnar valence-bond crystals
(VBC), and quantum spin liquids (QSL), with and without a spin gap, however no
consensus has been reached.

Materials, which may be proximate to the J1J2HM include Ba$_2$CuWO$_6$
\cite{Todate2007}, Sr$_2$CuMoO$_6$ \cite{Vasala2014a}, Sr$_2$CuWO$_6$
\cite{Vasala2014a, Vasala2014b}, Sr$_2$CuTeO$_6$ \cite{Koga2016}, and
Li$_2$VO(Si,Ge)O$_4$ \cite{Melzi2000, Melzi2001}.  These materials cover a wide
range of $\kappa$-values, realizing both N\'eel- and collinear ordered states.
Unfortunately, a system in the most frustrated region, $\kappa \sim 0.5$, is
still missing.

Apart from investigating the single layer case, it is of relevance, to extend the
parameter space of spin systems by introducing further interactions. These can
introduce additional well defined quantum phases, the connection of which to the
single layer case can provide for more insight.
Most popular along this line is the replication of a spin system in terms of 
antiferromagnetic dimers, forming, e.g., ladders \cite{Elbio1996}, bilayers 
\cite{Wang2006}, and three-dimensional networks \cite{Matsumoto2004}.
For strong dimer exchange $J_\perp$, these systems display a near product-state of 
weakly coupled singlets, the quantum dimer (QDM) phase, which features massive 
triplet excitations (triplons).
While in some cases the reduction of the dimer exchange may lead to condensation of 
triplons into sought-for quantum phases of some non-dimerized original model, 
bilayer systems host their own unique set of physics.
For this reason, a variety of bilayer systems have previously been studied under 
various objectives, e.g., uncovering rich phase diagrams \cite{Zhang2016, 
Zhang2018, Seifert2018, Joshi2019, Acevedo2021}, examining the crossing to 3D bulk 
materials \cite{Szalowski2012}, investigating effects of disorder 
\cite{Hoermann2020} and hole doping \cite{Nyhegn2022} or analyzing emergent bound 
states \cite{Wagner2021} and topological excitations \cite{Ghader2021}.
On the material side, an extensive number of systems exist, which are related to 
this theme, including Li$_2$VOSiO$_4$ \cite{Melzi2001}, BaCuSi$_2$O$_6$ 
\cite{Sebastian2006, Allenspach2021}, TlCuCl$_3$ \cite{Merchant2014}, 
Ba$_3$Mn$_2$O$_8$ \cite{Stone2008}, and SrCu$_2$(BO$_3$)$_2$\cite{Kageyama1999}.

The dimer version of the J1J2HM, forming an AA-stacked bilayer (J1J2BHM), has
been considered by modified spin-wave theory \cite{Hida1996}, which results in
magnetic order for all $\kappa$ as $J_\perp\rightarrow 0$, by dimer series
expansion \cite{Hida1998} for the spin-gap and the staggered susceptibility,
and recently by application of the high-order coupled cluster method (CCM) 
\cite{Bishop2019} for the magnetization. The latter two studies find a 
pa\-ra\-mag\-ne\-tic region for $J_\perp\rightarrow 0$ in the range $0.45\lesssim 
\kappa \lesssim 0.65$ and $0.43 \lesssim \kappa \lesssim 0.61$, respectively. This 
is consistent with studies of the single layer J1J2HM.

Dynamical properties, e.g., one- and two-triplon excitations in the QDM phase
of the J1J2BHM, as well as their condensation into the ordered phases versus
$\kappa$ and $J_\perp$ remain open issues. This provides the main motivation
for our work. We will analyze the spectrum of the J1J2BHM up to the two-triplon
sector, starting from the limit of decoupled dimers. We will use the
perturbative Continuous Unitary Transformation (pCUT) \cite{Knetter2000a},
based on the flow equation method \cite{Wegner1994}, in order to perform a
series expansion for the excitation energies directly in the thermodynamic limit. 
pCUT has been applied successfully to a large variety of dimerized and n-merized 
quantum spin systems, including, but not limited to ladders \cite{Windt2001}, tubes
\cite{Arlego2013}, planar pyrochlores \cite{Brenig2002}, various SU(2)-invariant
Heisenberg bilayers \cite{Zhang2016, Hoermann2020}, as well as to Kitaev
bilayers \cite{Seifert2018, Wagner2021}.

The paper is organized as follows:  In Sec.~\ref{sec:model} the J1J2BHM is
described. Sec.~\ref{sec:method} provides for a general explanation of the pCUT
method.  Sec.~\ref{sec:results} details our results, i.e., the one-triplon
excitations in Sec.~\ref{sec:results:one} and two-triplon excitations 
(Sec.~\ref{sec:results:two}). Sec.~\ref{sec:conclusion} concludes our work and lists
some speculations. A technical appendix, Sec.~\ref{sec:appendix:euler} , on
specifics of a resummation method we use is included.

\section{Model}
\label{sec:model}

\begin{figure}[tb]
	\includegraphics[width=0.9\columnwidth]{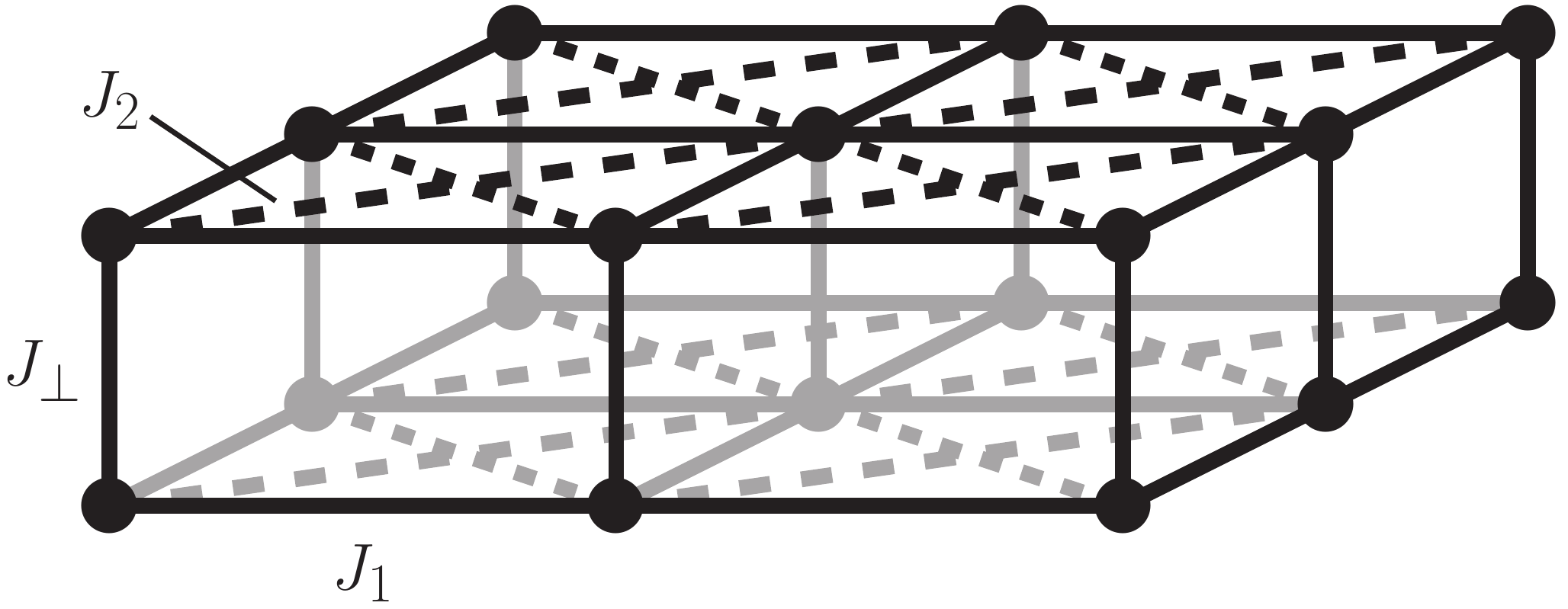}
	\caption{The $J_1$-$J_2$-Heisenberg square lattice bilayer. Each $\bullet$
		hosts a spin-\sfrac{1}{2}.}
	\label{fig:1}
\end{figure}

The Hamiltonian of the J1J2BHM reads
\begin{equation}
	\begin{aligned}
		H &= H_0 + H_I \\
		H_0 &= J_\perp \sum_{\mathbf{r}} \vec{S}_{\mathbf{r},1}\cdot 
		\vec{S}_{\mathbf{r},2} \\
		H_I &= J_1 
		\sum_{\mathclap{\substack{\langle\mathbf{r},\mathbf{r}'\rangle
					\\ L = 1,2}}} 
		\vec{S}_{\mathbf{r},L}\cdot \vec{S}_{\mathbf{r}',L} + J_2 
		\sum_{\mathclap{\substack{\langle\langle\mathbf{r},\mathbf{r}' 
		\rangle\rangle \\ L = 1,2}}} 
		\vec{S}_{\mathbf{r},L}\cdot \vec{S}_{\mathbf{r}',L}
	\end{aligned}
\end{equation}
where $\vec{S} = \lbrace S^\alpha\rbrace$ with $\alpha = x,y,z$ are 
spin-\sfrac{1}{2} operators, $J_\perp$ and $J_{1(2)}$ are the Heisenberg 
interlayer and (next-) nearest neighbor intralayer exchange, respectively.
$L = 1,2$ labels the two layers, $\mathbf{r}^{(\prime)}$ the sites of the 
square lattice and $\langle\mathbf{r},\mathbf{r}'\rangle$, 
$\langle\langle\mathbf{r},\mathbf{r}'\rangle\rangle$ denote NN and NNN sites.
Fig.~\ref{fig:1} shows a depiction of the sites and spin exchanges.

In general the couplings can be any combination of ferro- or antiferromagnetic 
interactions.
In this study we only focus on the pure antiferromagnetic case, i.e., 
$J_\perp, 
J_1, J_2 > 0$.
Further $J_\perp \equiv 1$ is chosen from here onwards to fix the energy scale 
and we will use the parameter $\kappa = \frac{J_2}{J_1}$ to measure the 
strength of the NNN-interactions.

\section{Method}
\label{sec:method}

The J1J2BHM under study features at least three limiting quantum phases which 
are adiabatically disjoint.
First for $J_\perp \gg J_1, J_2$ the system can be viewed as weakly interacting 
antiferromagnetic dimers and serves as the starting point for our studies. 
Second, in the case $J_1 \gg J_\perp, J_2$, N\'eel order on each layer is 
present.
In a similar manner, for $J_2 \gg J_\perp, J_1$, each layer realizes a collinear
magnetic order, i.e., a N\'eel order on each bipartite sublattice, while the 
relative angle between the spins on both sublattices is fixed through 
order-by-disorder selection.

We study the J1J2BHM starting from the limit $J_\perp \gg J_1, J_2$, following 
the same evaluation scheme as in earlier work on the Kitaev-Heisenberg-Bilayer 
\cite{Seifert2018, Wagner2021}.
For completeness we reiterate the main points:

The non-degenerate ground state of the model is formed as a product state of 
singlets on each dimer.
The corresponding elementary excitations, i.e., triplets excited on single 
dimers, can be classified by their $S^z$-component.
We write
\begin{equation}
	\begin{aligned}
		\ket{t_{+1}} &= \ket{\uparrow\uparrow} \\
		\ket{t_0} &= \left(\ket{\uparrow\uparrow} + 
		\ket{\downarrow\downarrow}\right)/\sqrt{2} \\
		\ket{t_{-1}} &= \ket{\downarrow\downarrow}
	\end{aligned}
\end{equation}
and refer to them as $\alpha = +1, 0, -1$-triplons hereafter.
These properties allow us to apply the perturbative Continuous Unitary 
Transformation (pCUT) technique \cite{Knetter2000a}, based on the flow-equation 
method \cite{Wegner1994}, to the model, requiring the unperturbed Hamiltonian 
$H_0$ to have a non-degenerate ground state and an equidistant spectrum.
In the present case each energy level of the spectrum of $H_0$ can be assigned 
a particle number $Q \geq 0$, i.e., the number of excited triplets, which 
describes the energy of the unperturbed states, i.e., $H_0~=~J_\perp Q\,+\, 
\text{const.}$ and especially $[H_0, Q] = 0$.
$Q = 0$ refers to the product ground state $\ket{} = \prod_{\mathbf{r}} 
\ket{s_\mathbf{r}}$ of singlets, while the one- and two-triplon states, i.e., 
$Q = 1$ and $Q = 2$, are writen as $\ket{\mathbf{r} \alpha} = 
\ket{t_{\mathbf{r},\alpha}} \otimes \prod_{\mathbf{r'} \neq \mathbf{r}} 
\ket{s_{\mathbf{r}'}}$ and $\ket{\mathbf{r} \alpha, \mathbf{r}' \beta} = 
\ket{t_{\mathbf{r},\alpha}} \otimes \ket{t_{\mathbf{r}',\beta}} \otimes 
\prod_{\mathbf{r}'' \neq \mathbf{r},\mathbf{r}'} \ket{s_{\mathbf{r}''}}$, 
respectively.

The perturbation $H_I$ of the Hamiltonian mixes different $Q$-sectors through
creating or destructing triplons.
By virtue of pCUT, the full Hamiltonian $H$ is transformed to an effective 
Hamiltonian $H_{\text{eff}} = U H U^\dagger$, which is $Q$-diagonal and can be 
expressed by a series in the perturbation parameters $J_{1(2)}$ as
\begin{equation}
	H_{\text{eff}} = H_0 + \sum_{l,m}^{\infty} C_{l,m} J_1^l J_2^m\ ,
\end{equation}
where $C_{l,m}$ are weighted products of terms in $H_I$, each comprising $l + 
m$ non-local creations(destructions) of triplons which in total conserve the 
$Q$-number.
The weights of the $C_{l,m}$ are integer fractions, which are determined 
analytically, and independent of the specific model at hand, by recursive 
differential equations \cite{Knetter2000a}.
Due to transforming the Hamiltonian as a whole, pCUT works directly in the 
thermodynamic limit, and due to its perturbative nature, the expansion is exact up 
to the order calculated. Thus its results are well controlled for small parameters 
$J_{1(2)}$.

Using the $Q$-number conservation, we evaluate the spectrum by treating 
each sector independently.
For this we determine the irreducible matrix elements of $H_{\text{eff}}$ for 
each value of $Q = 0,1,2$ and solve the corresponding zero-, one- and 
two-particle problems, described in the following:

Due to the uniqueness of the ground state, the $Q = 0$-case is described by a
single matrix element, directly equaling the ground state energy $E_0 = 
\bra{}H_{\text{eff}}\ket{}$.

The one-particle case is described by a translational-invariant matrix, leading 
to the one-particle dispersion $\mathbf{E}_{\mathbf{k},\alpha\beta} = 
\sum_{\mathbf{r}} \mathrm{e}^{i\mathbf{r}\mathbf{k}} \bra{\mathrm{r} 
\alpha}H_{\text{eff}}\ket{\mathbf{0} \beta} - \delta_{\mathbf{r},\mathbf{0}} 
\delta_{\alpha\beta} E_0^{\text{cl}}$, where $\alpha,\beta \in \lbrace 
+1,0,-1\rbrace$, $\mathbf{k}$ is a wavevector and $E_0^{\text{cl}}$ is the 
ground state energy calculated for the same cluster as the corresponding 
one-particle matrix element \cite{Knetter2000a, Knetter2003t}.
In general this is a $3 \times 3$-matrix, however, due to the $SU(2)$-symmetry 
of the model, the total $z$-component of the spins has to be conserved and thus 
different triplon flavors do not mix in the one-particle sector.
In fact, they must have identical dispersions, we write 
$\mathbf{E}_{\mathbf{k},\alpha\beta} \equiv E(\mathbf{k}) 
\delta_{\alpha\beta}$ and only refer to the dispersion $E(\mathbf{k})$ in the 
following.
To check at least part of our series coefficients, we can compare the case 
$J_2 = 0$, i.e., the non-frustrated version of the bilayer, to previous results 
from the literature \cite{Zheng1997}.

The two-particle problem, i.e., $Q = 2$, is more challenging 
\cite{Knetter2003t, Knetter2003}.
Here the matrix elements $\bra{\mathbf{r}' \alpha', \mathbf{r}'+\mathbf{d}' 
\beta'}H_{\text{eff}}\ket{\mathbf{r} \alpha, \mathbf{r}+\mathbf{d} \beta}$ 
describe two particles with initial(final) positions $\mathbf{r}^{(\prime)}$ 
and $\mathbf{r}^{(\prime)}+\mathbf{d}^{(\prime)}$ and triplon flavors 
$\alpha^{(\prime)}$, $\beta^{(\prime)}$.
Similar to the one-particle case, an effective two-triplon Hamiltonian 
{\itshape matrix} $h_{\mathbf{K}}(\mathbf{d}, \mathbf{d}', \alpha\beta, 
\alpha'\beta')$ with respect to states $\ket{\mathbf{K}, \mathbf{d}, 
\alpha\beta}$ can be constructed, where $\mathbf{K}$ is the total momentum and 
$\mathbf{d}$ labels the two-triplon separation.
This matrix $h_{\mathbf{K}}$, comprising of the analytical matrix elements of 
the SE, 
has a particular structure, directly representing the underlying physics, i.e., 
the two-particle scattering problem, that can be used to extract the 
two-triplon spectrum.
Most important, $h_{\mathbf{K}}$ is band-diagonal with respect to $\mathbf{d}$, 
due the model only involving local spin interactions, and comprises of two 
different types of matrix elements:
First, for $|\mathbf{d}^{(\prime)}| < d_I$, i.e., a two-triplon separation 
smaller then some characteristic length $d_I$, both triplons interact 
through an effective coupling determined by pCUT and the corresponding matrix 
elements describe those irreducible two-triplon interactions.
Second, for larger $\mathbf{d}^{(\prime)}$, the triplons are too far separated 
to interact, but can still move separately across the lattice.
This forms a semi-infinite band in $h_{\mathbf{K}}$, describing the propagation 
of scattering states.
The latter allows us to diagonalize $h_{\mathbf{K}}$ numerically on 
sufficiently large lattices with periodic boundary conditions without 
neglecting any two-particle interactions.
Thus the resulting spectrum will capture all relevant two-triplon states, 
especially any (anti-)bound states outside the two-particle continuum.
In practice we evaluate the spectrum of $h_{\mathbf{K}}$ on a system with $20 
\times 20$-dimers (800 spins).
\\
Again, the $SU(2)$-invariance of the model fixes the total $z$-component of a 
two-triplon state, resulting in a block-diagonal form of $h_{\mathbf{K}}$ under 
the constraint $\alpha + \beta = \alpha' + \beta'$.
More over, the total spin $S = 0,1,2$ of the two-triplon state is conserved as 
well, providing us with a suitable classification for the spin-structure 
of the two-triplon states, we write $\ket{\mathbf{K},\mathbf{d},\alpha\beta} 
\to \ket{\mathbf{K},\mathbf{d},S,S^z}$.
In practice, this is used to simplify the evaluation of the SE, while also 
providing an additional check for the resulting states after the numerical 
diagonalization.

All evaluations of required matrix elements of $H_{\text{eff}}$ can be carried 
out on suitable chosen linked cluster graphs of the lattice.
A detailed description of their construction procedure can be found in Ref. 
\cite{Wagner2021}.

\section{Results}
\label{sec:results}

In this section we describe our findings on the low energy spectrum of the 
J1J2BHM.
Sec.~\ref{sec:results:one} covers one-particle excitations, while 
Sec.~\ref{sec:results:two} contains the results for the two-particle states.
In both cases, we investigate the structure of the spectrum as well as the 
wave function of the excitations and determine a presumed outline of the 
phasediagram.

\subsection{One-particle excitations}
\label{sec:results:one}

\begin{figure}[tb]
	\includegraphics[width=\columnwidth]{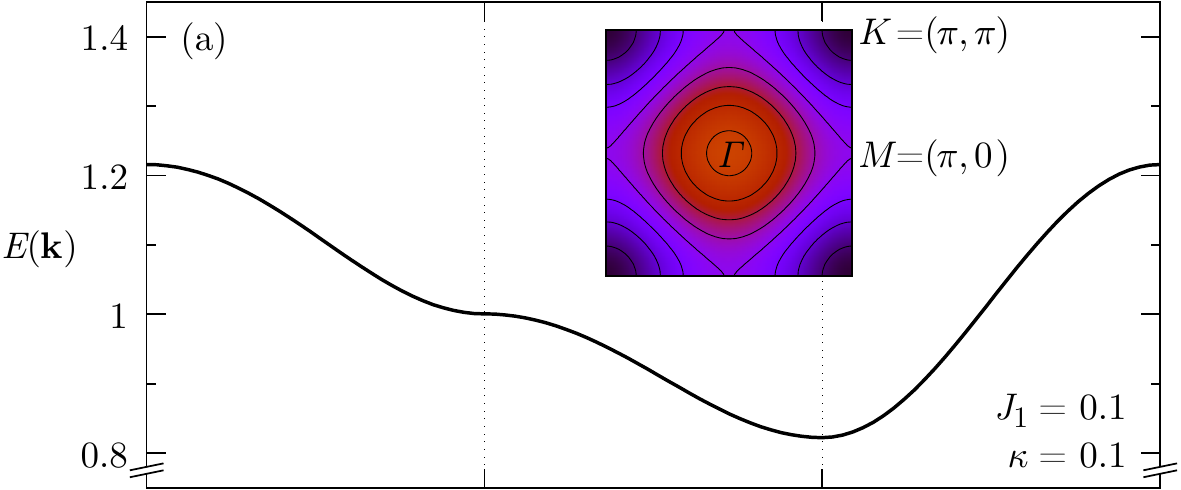}
	\includegraphics[width=\columnwidth]{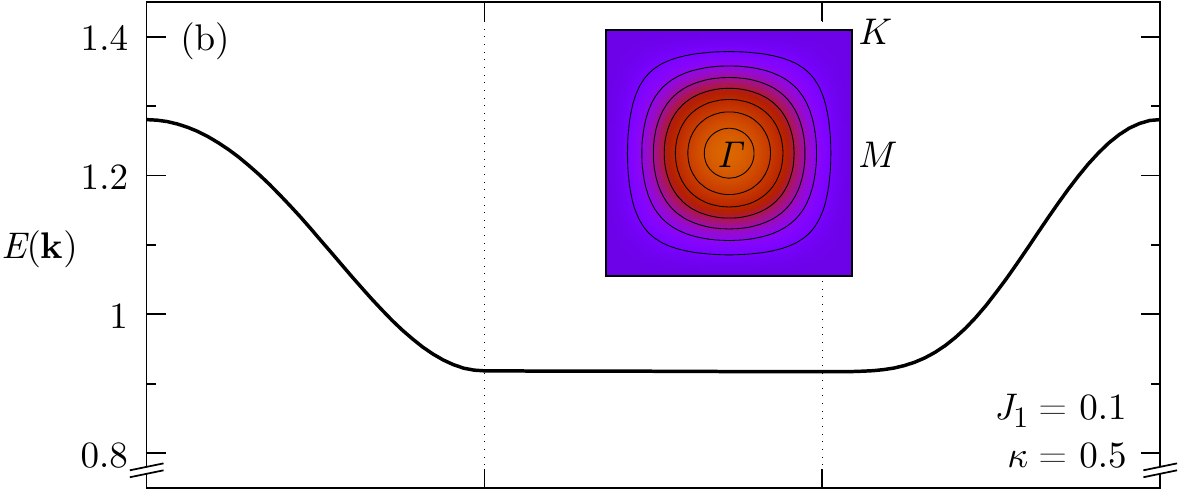}
	\includegraphics[width=\columnwidth]{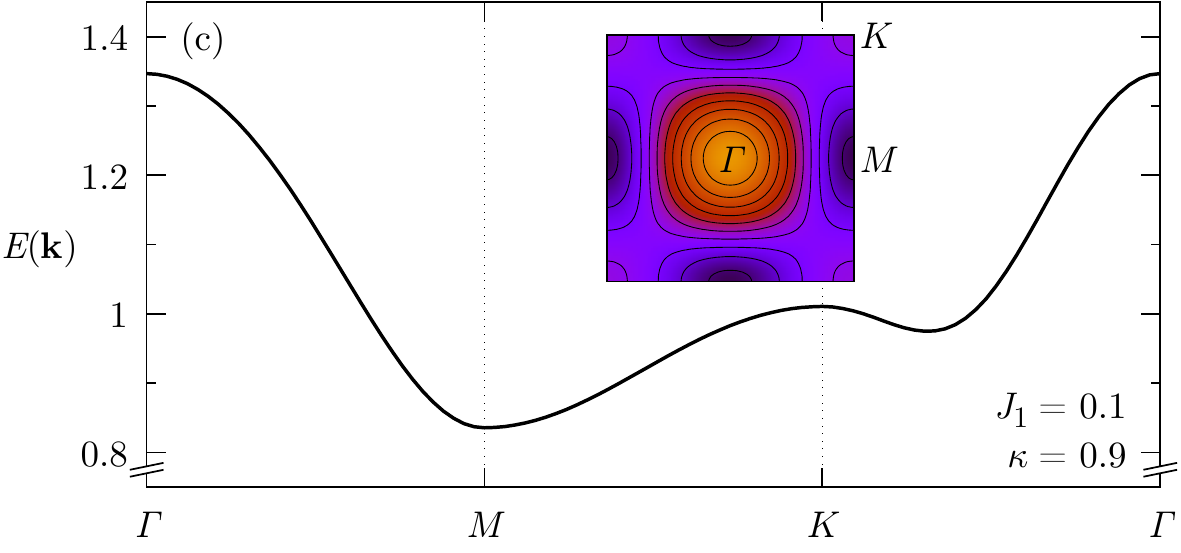}
	\caption{One-particle dispersion $E(\mathbf{k})$ in 7th order series 
		expansion for $J_1 = 0.1$ at varying values of $\kappa = 0.1, 0.5, 0.9$ 
		(top to bottom) along high-symmetry lines of BZ. Insets: constant 
		energy surfaces.}
	\label{fig:2}
\end{figure}

Fig.~\ref{fig:2} shows the one-triplon dispersion $E(\mathbf{k})$ along some 
symmetry lines of the Brillouin zone (BZ) for different values of $\kappa = 
\frac{J_2}{J_1}$ at a fixed value $J_1 = 0.1$, calculated to 7th order in the 
series expansion using pCUT.
From this, the overall tendencies of the model can already be visualized:

First, for $\kappa \ll 0.5$, the dispersion has a clear minimum at $\mathbf{k} =
(\pm\pi,\pm\pi)$, which coincides directly with the formation of N\'eel order in
the single layer model for small NNN coupling $J_2$.  Second, for $\kappa \gg
0.5$, the global minimum can be found at $\mathbf{k} = (\pi,0)$ (and its
equivalent points).  This fits the expectation from the single layer model, that
for sizeable $J_2$ a collinear magnetic order ground state is formed.  In both
cases the minima remain at their respective $\mathbf{k}$-points even for higher
$J_1$ to the point when the excitation gap vanishes.  Last, for $\kappa \approx
0.5$, the dispersion lacks a clear minimum.  Instead, almost all wave vectors
along the edge of the BZ acquire nearly identical energies. This is consistent
with the behavior of the single layer model, where at the maximally frustrated
point, a line-degeneracy at zero energy of the dispersion indicates the absence
of a well defined ordering vector. In the present case this scenario is
accompanied by the dimer gap.

\begin{figure}[tb]
	\includegraphics[width=\columnwidth]{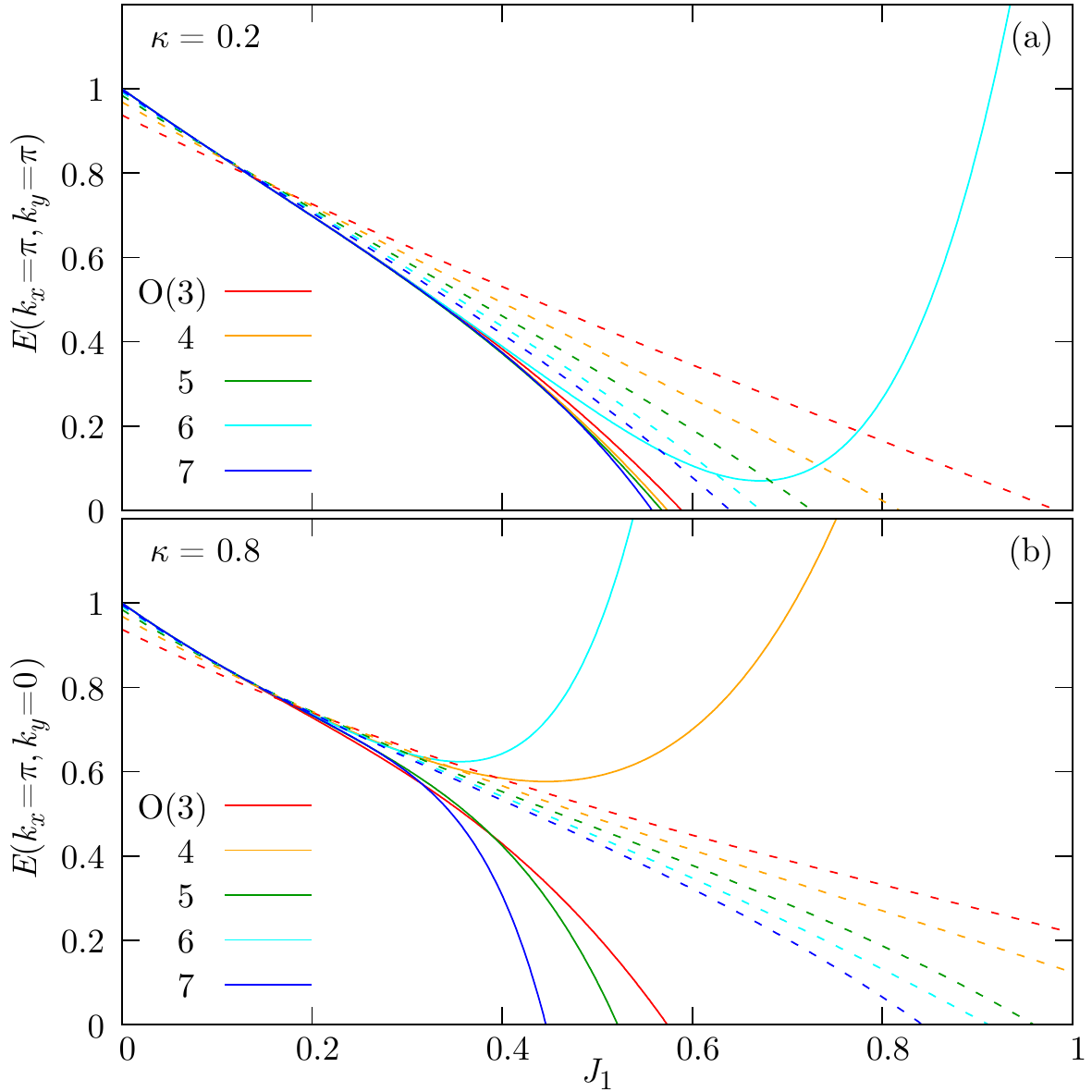}
	\caption{Comparison between bare series and series after Euler resummation 
		for orders 3 to 7 at (a) $\kappa = 0.2$ and $\mathbf{k} = (\pi,\pi)$ and 
		(b) $\kappa = 0.8$ and $\mathbf{k} = (\pi,0)$ versus NN coupling $J_1$.}
	\label{fig:3}
\end{figure}

Based on those observations, it is natural to investigate the excitation gap at
the most likely critical points $\mathbf{k} = (\pi,\pi)$ and $\mathbf{k} =
(\pi,0)$ with respect to the interaction strengths and varying orders of the
series expansion.  Fig.~\ref{fig:3}~(a) displays the triplon energy at $\mathbf{k}
= (\pi,\pi)$ in the N\'eel-near parameter space for $\kappa = 0.2$ and varying
$J_1$.  It shows a comparison between 3rd to 7th order bare series expansion and
equivalent results after an Euler resummation scheme is applied (details can be
found in appendix \ref{sec:appendix:euler}).  In this case, it appears that the
convergence of the series is satisfying, even without the application of a
resummation scheme for most orders investigated. Only the series at $O(6)$ is an
outlier, diverging in the vicinity of the tentative critical point.  The
resummation corrects this and leads to a monotonic convergence of the critical
point.  We find $J_{1,c} \approx 0.64$ for $\kappa = 0.2$ at our highest
expansion order using the resummation.

For larger $\kappa$ and at $\mathbf{k} = (\pi,0)$ the situation differs, see
Fig.~\ref{fig:3}~(b).  Here, we observe an alternating behavior between even and odd
orders of the bare series expansion, with odd orders providing a gap closure,
while even orders diverge. Applying Euler resummation as for the case of
$\kappa<0.5$, this issue can be resolved, again leading to a monotonic decrease
of the critical coupling with increasing expansion order.  Using resummation, we
find $J_{1,c} \approx 0.84$ for $\kappa = 0.8$ at the largest expansion order.

\begin{figure}[tb]
	\includegraphics[width=\columnwidth]{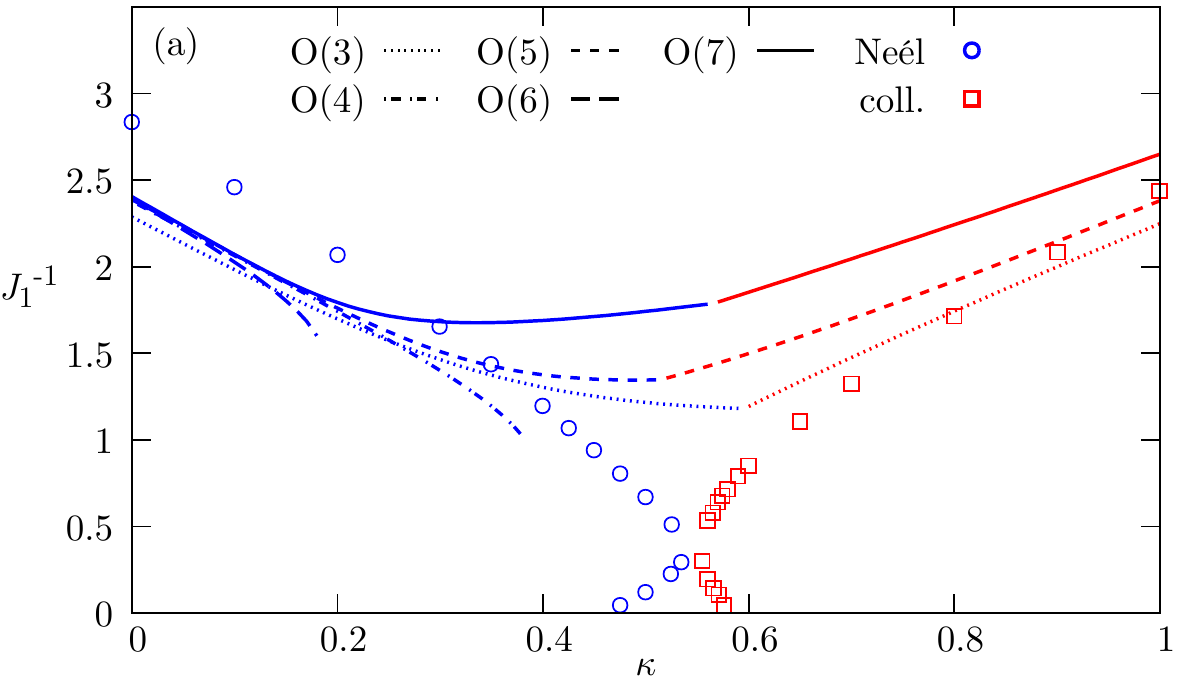}
	\includegraphics[width=\columnwidth]{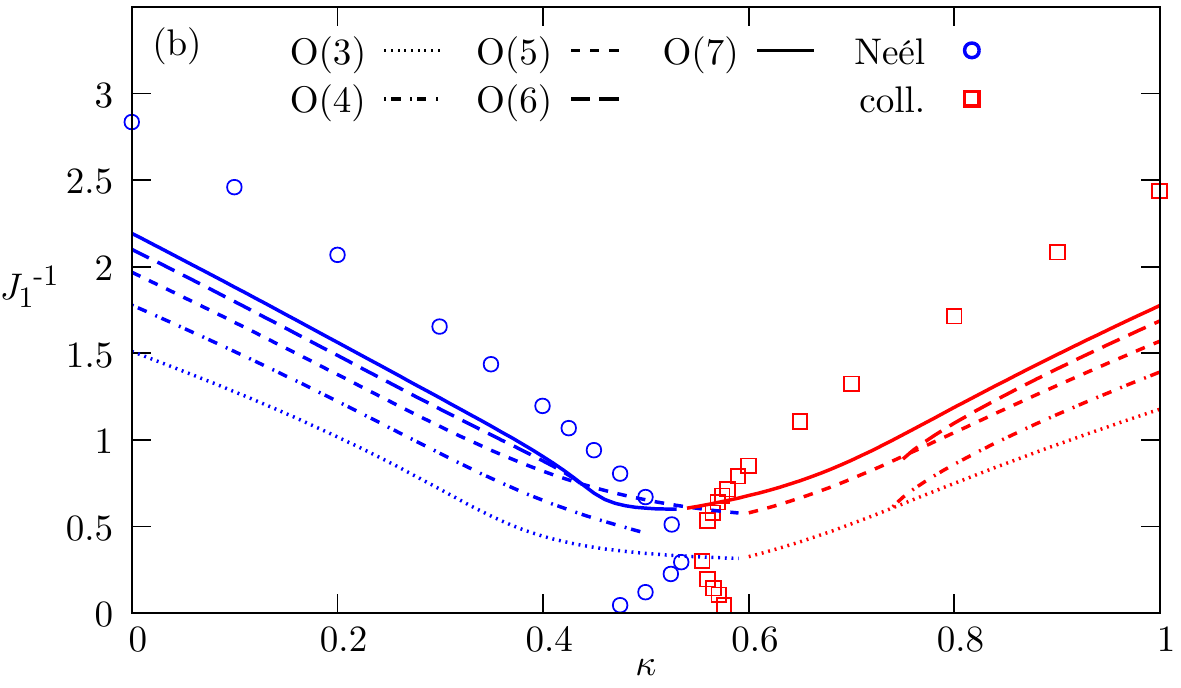}
	\caption{Phase boundaries as determined through the one-triplon gap 
		closure for varying orders in (a) bare series expansion and (b) Euler 
		resummation. Open dots and squares show results from Ref. 
		\cite{Bishop2019}. Color shows the wavevector of the one-triplon gap 
		closure.}
	\label{fig:4}
\end{figure}

Next, we use such findings to determine the phase diagram for the J1J2BHM.
Because our approach provides the complete triplon dispersion as power series in
$J_1$ and $\kappa$, we can easily scan for the critical coupling
$J_{1,c}(\kappa)$ at which the triplon energy vanishes for a given wavevector
for any value of $\kappa$. I.e., similar to the preceding paragraphs, we now
consider the gap closings at $\mathbf{k} = (\pi,\pi)$ (N\'eel-type) and
$\mathbf{k} = (\pi,0)$ (collinear-type) versus $\kappa$ and plot the resulting
phase boundaries for various orders in Fig.~\ref{fig:4}. Panel (a) shows
the results of the bare series expansion, while panel (b) shows those for the
resummed series. For the two pitch vectors considered, in both cases only those
boundaries are shown for which the excitation gap closes first.

Several remarks are in order: First, the phase diagram is displayed in terms of
the parameters $(\kappa, \frac{1}{J_1})$. This allows for direct comparison of
our findings to those from CCM in Ref. \cite{Bishop2019}. The latter are marked
by open dots and squares in the figure. In terms of this parameter space, the
decoupled dimer limit corresponds to regions of large $\frac{1}{J_1}$, while the
single layer model resides on the lower x-axis at $\frac{1}{J_1} = 0$.  Second,
contrasting panel (a) against (b) the influence of the series resummation is
apparent. While in (a), and in particular for even orders, there are large
windows of $\kappa$ with no gap-closing, for (b) these windows shrink. Moreover,
the evolution versus expansion order in (b) suggest a well behaved convergence,
with $O(7)$ critical lines from resummation not too far from the infinite order
limit. In contrast to the CCM \cite{Bishop2019}, our results from the
one-triplon gap predict a slightly more extended dimer phase for small
$\kappa$ based on the bare series and for all $\kappa$ based on the resummed
series. This variance likely stems from a bias imposed on the two methods by
virtue of their opposite "starting phases", i.e., dimer (LRO) phase for the
pCUT (CCM).

For all $\kappa$ and in particular in the region of maximal frustration near
$\kappa \approx 0.5$ the critical lines of neither the bare nor the resummed
series show a tendency to approach $\frac{1}{J_1} \to 0$ upon increasing the
expansion order. Rather, as one can see from Fig.~\ref{fig:4}~(b), there is a
clear tendency near $\kappa\approx 0.5$, to stabilize the critical line for
single triplon gap-closure at some minimum finite value of $\frac{1}{J_1} \sim 
0.65$. This is remarkably close to the termination of reentrant behavior of the 
N\'eel and collinear phases, observed by CCM \cite{Bishop2019}. From the latter, 
the non-magnetic and potentially spin-liquid regime of the single-layer model, for 
$0.45 \lesssim \kappa \lesssim 0.59$, extends upwards, forming an "hourglass" 
shaped region at finite $J_\perp$, visible in Fig.~\ref{fig:4}~(b). Combining this 
with the critical line from the resummed SE, it is very tempting to speculate that 
the single-layer QSL, anticipated on the line $\frac{1}{J_1}=0$, is confined to the 
lower part of this hourglass and terminates within its constriction. This is very 
reminiscent of a somewhat similar situation of a QSL surrounded by a QDM and two 
reentrant LRO phases in the frustrated honeycomb bilayer Heisenberg model 
\cite{Zhang2018}.

\subsection{Two-particle excitations}
\label{sec:results:two}

\begin{figure}[tb]
	\includegraphics[width=\columnwidth]{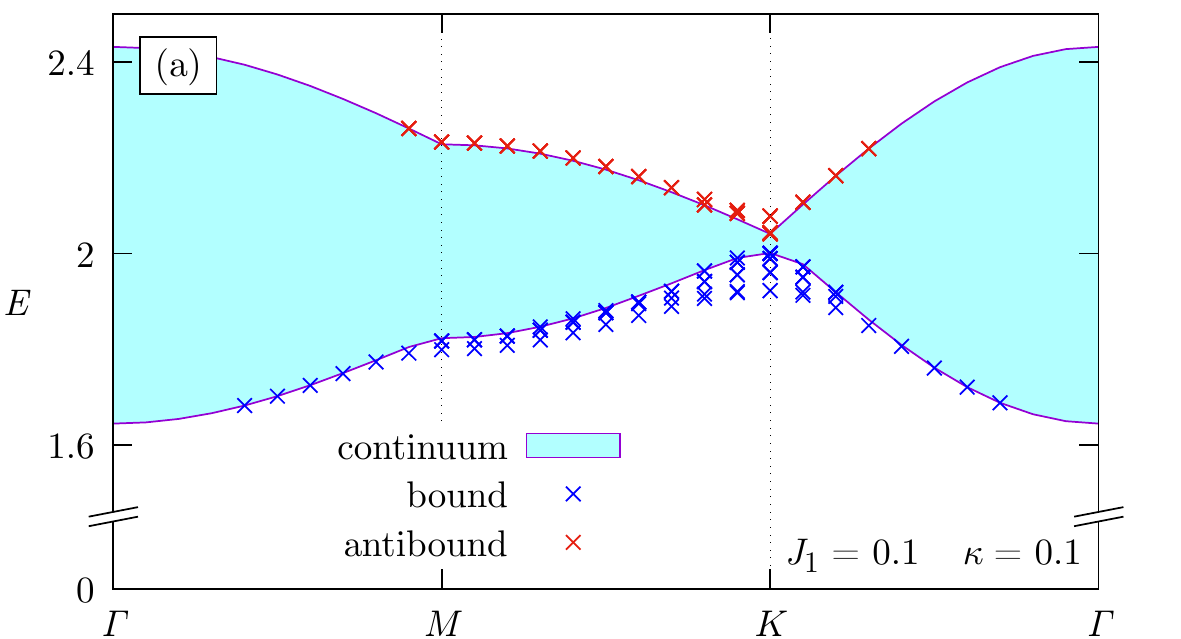}
	\includegraphics[width=\columnwidth]{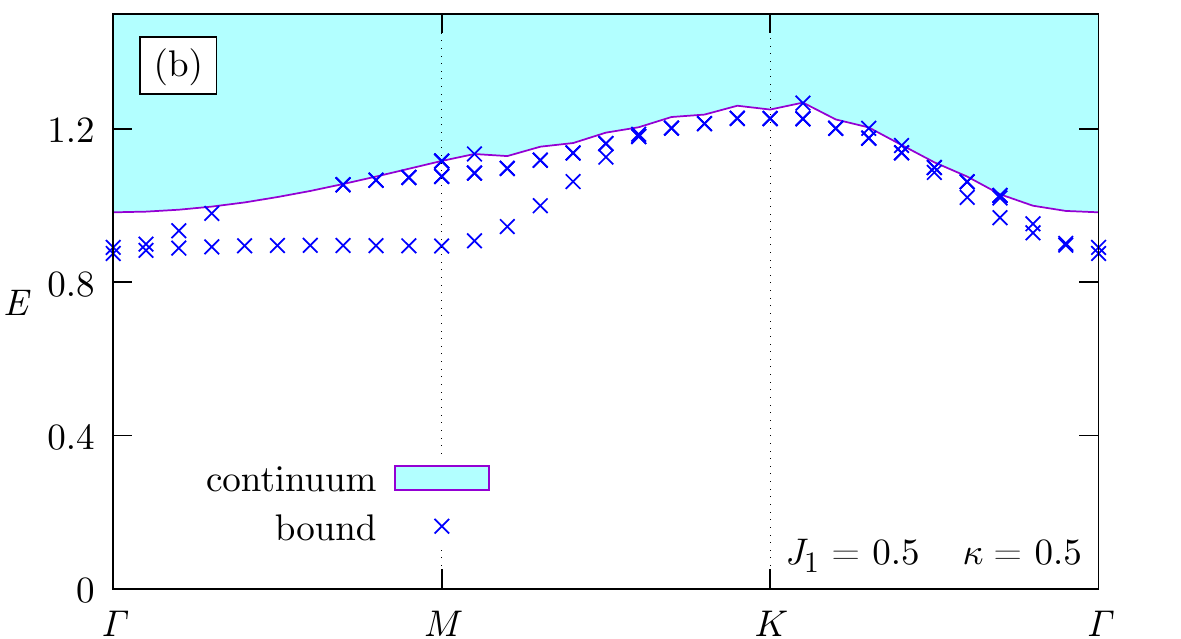}
	\includegraphics[width=\columnwidth]{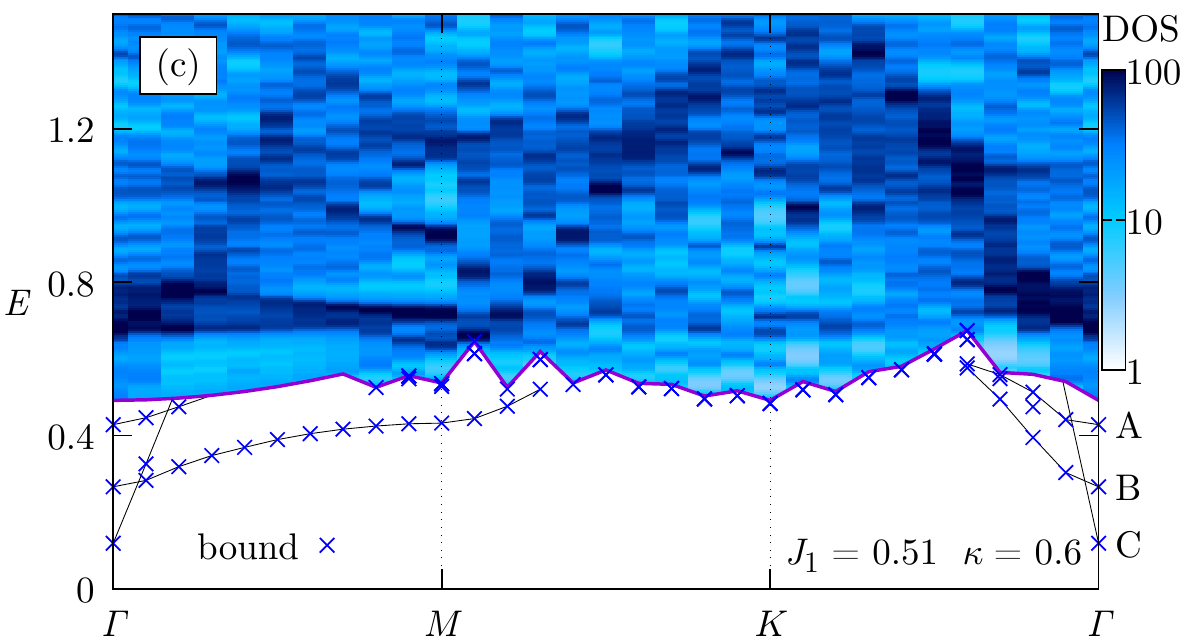}
	\caption{Two particle spectra for three different parameter sets over 	
		total momentum $\vec{K}$ along various high-symmetry lines in the BZ for 
		series expansion at 7th order (one-particle matrix elements) and 5th order 
		(two-particle interactions). Panel (c): DOS of continuum indicated by color 
		scale (in arbitrary units). Thin lines are guides to the eye for states 
		with similar wave functions. A, B, C label bound states depicted in 
		Fig.~\ref{fig:6}.}
	\label{fig:5}
\end{figure}

In this section we focus on the two-triplon excitations. Fig.~\ref{fig:5} shows
their spectrum versus the total momentum $\mathbf{K}$ along selected
high-symmetry paths in the BZ for different values of $J_1$ and $\kappa$.  A few
notes are in order.  First, for each total momentum $\mathbf{K}$, the spectrum
comprises of two parts, i.e., a continuum of states and potentially several
discrete (anti-)bound states.  The continuum is formed from all combinations of
two one-triplon states with energies $E(\mathbf{k}_1)$ and $E(\mathbf{k}_2)$ and
total momentum $\mathbf{K} = \mathbf{k}_1 + \mathbf{k}_2$.  The (anti-)bound
states can in principle occur at any $\mathbf{K}$. Their formation and size of
splitting from the continuum however, i.e., (anti)binding energy, depends
strongly on the specifics of the two-triplon scattering potential and the total
wavevector. In particular, for small $\kappa$, we find (anti-)bound states
primarily in the vicinity of $\mathbf{K} = (\pm\pi,\pm\pi)$, where the
two-triplon continuum is narrowest, see Fig.~\ref{fig:5}~(a). Satisfyingly, these
states are consistent with previous results from a different type of series
expansion performed at $\kappa = 0$ only, i.e., the $J_1$-Heisenberg-Bilayer
\cite{Collins2008}.

Increasing $\kappa$, the width of the continuum in the vicinity of the
$K$-point increases and the dominant (anti)bound states start to move. For
intermediate $\kappa$ at the lower edge of the region of strongest frustration,
i.e., $\kappa\sim 0.5$ and Fig.~\ref{fig:5}~(b), well split-off, low-energy
bound states can be found all along $\Gamma$-$M$. For even larger $\kappa$,
i.e., at the upper edge of the most frustrated region, Fig.~\ref{fig:5}~(c),
bound states at the $\Gamma$-point are lowest in energy. Most remarkably, here,
we find these latter energies to be lower than {\em all} of those from the
complete one- {\em and} two-triplon spectrum. This renders the two-particle bound 
states the low-energy elementary excitations of the system. We will elaborate on 
this later.

Two additional features are visible in Fig.~\ref{fig:5}~(c). First, for larger
$\kappa$, small ripples at the boundary of the continuum occur. This is {\em
not} due to the series being of insufficiently high order, rather it is a finite
size effect of particular nature. I.e., for any finite size of the one-triplon
momentum space, the actual total momentum of the lower edge of the two-triplon
continuum may be off from the available momenta $\mathbf{k}_1 + \mathbf{k}_2$,
leading to the ripples visible. Second, and interestingly, a contour plot of the
continuum two-triplon density of states (DOS) in Fig.~\ref{fig:5}~(c) suggests, 
that a qualitative understanding of the positions of the bound states can be drawn 
from a T-matrix type of argument. I.e., in regions of high two-particle DOS close 
to the lower edge of the continuum, along $\Gamma$-$M$, bound states are 'pushed' 
away from the continuum, while closer to the $K$-point, large two-particle DOS is 
observed only further into the continuum, leaving the bound states very close to the
continuum boundary.

\begin{figure}[tb]
	\includegraphics[width=\columnwidth]{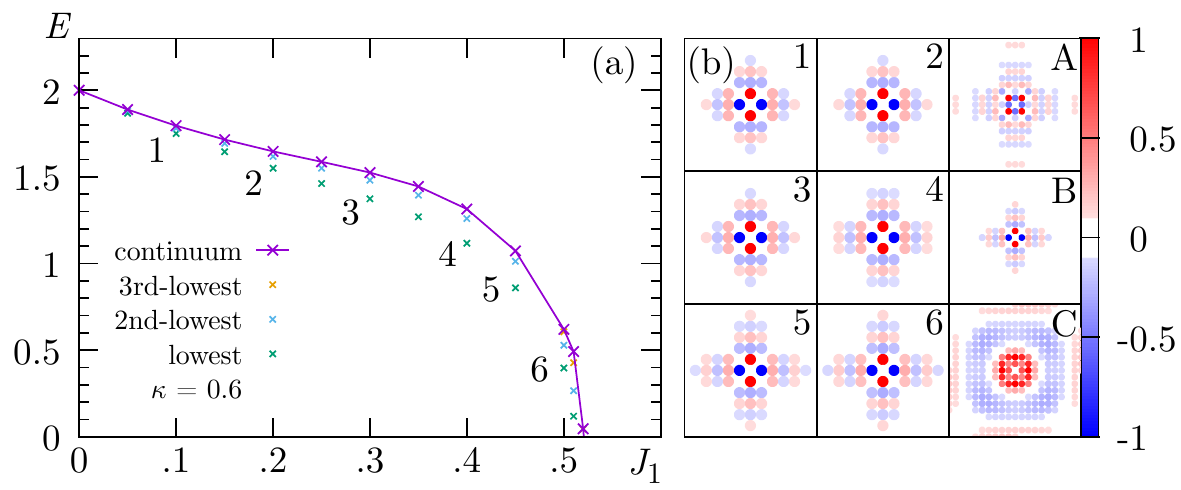}
	\caption{ (a) Energies of the bound states and lower edge of the 
		two-triplon continuum at $\vec{K} = (0,0)$ versus $J_1$ at $\kappa = 0.6$.  
		(b) Density plots of the wave functions of the lowest energy bound states 
		versus $J_1$ from panel (a), marked 1 through 6, as well as all bound 
		states at $J_1 = 0.51$, marked A through C, compare Fig.~\ref{fig:5} (c), 
		on the planar square-lattice grid of two-triplon separations $\mathbf{d}$.  
		Amplitudes $C_{\mathbf{d}}$ are renormalized to $C_{\mathbf{d}} \in [-1,1]$ 
		for each state and only those with $\left\vert C_{\mathbf{d}} \right\vert^2 
		> 0.01$ are shown. Each plot is centered at $\mathbf{d} = (0,0)$.}
\label{fig:6}
\end{figure}

Now we turn to the internal structure of the bound states uncovered.  As
described in Sec.~\ref{sec:method}, the spin-component of the two-triplon
eigenstates of $H_{\text{eff}}$ can be classified according to total spin
$S=0,1,2$ and its $z$-component $S^z = -S,\cdots,+S$.  We find that all bound
states we obtain are $S=0$ and $S=1$ states, while the antibound states satisfy
$S=2$.  Apart from the spin-quantum number, the two-triplon wave function can be
classified according to lattice harmonics, i.e., the 'angular' momentum
loosely speaking. For that purpose we consider the $\mathbf{d}$-dependence of
the two-triplon wave function
\begin{equation}
	\ket{\Psi_n(\mathbf{K})} = \sum_{\mathbf{d}} C_{\mathbf{d}}
	\ket{\mathbf{K},\mathbf{d},S_n^{\phantom{z}},S_n^z}\ ,
\end{equation}
where the $C_{\mathbf{d}}$ are wave function amplitudes, which in principle are
complex. However, at $\mathbf{K} = (0,0)$, and because all exponentials of type
$\sim\exp(i{\bf K}(\dotsc))$ are unity, the effective Hamiltonian in each
$S,S^z,Q{=}2$-subspace is real and symmetric. In turn, the $C_{\mathbf{d}}$ are
real numbers at that point.  By construction, $C_{-\mathbf{d}} = C_{\mathbf{d}}$
due to the triplons being indistinguishable. This defines a basic symmetry for
all spatial wave functions considered. Morover, we can now identify additional
symmetries of the (anti)bound states, from the wave functions exemplified in
Fig.~\ref{fig:6} (b) for $\kappa = 0.6$ at $\mathbf{K} = (0,0)$.  Namely first,
there are states with the largest $C_{\mathbf{d}}$ for $\mathbf{d}$ along NN
$J_1$-bonds. These are the lowest energy bound states.  At $\mathbf{K} = (0,0)$
these states are sign reversed along the two orthogonal lattice directions,
e.g., $C_{(1,0)} = -C_{(0,1)}$, i.e., they are odd under rotation by $\pi/2$
(states '1' to '6' and 'B', Fig.~\ref{fig:6}~(b)).  Second, there are states
with largest amplitude along the NNN $J_2$-bonds (state 'A', Fig.~\ref{fig:6}~(b)). 
These states are invariant under rotation by $\pi/2$. Finally, states of
type 'C' in Fig.~\ref{fig:6}~(b) show a quasi-cylindrical symmetry. This seems
unnatural for the underlying square lattice.

Fig.~\ref{fig:6}~(a) highlights the evolution of the energies of the various
bound states versus $J_1$. The figure details, that for all $J_1 < 0.5$, the
lowest lying bound states display 'B'-symmetry, i.e., they are odd under
$\pi/2$ rotation and are tightly bound with a dominant NN-amplitude. Only for
$J_1 \gtrsim 0.5$ lowest-energy bound states of type 'C' (see lower right corner
of Fig.~\ref{fig:6}~(b)) emerges. We argue that the latter states are unphysical
artifacts of our SE. First, as can be seen in Fig.~\ref{fig:5}~(c), their
dispersion is unusually steep, with only very few $\mathbf{K}$-points outside of
the continuum. Second, their wave function is rather extended with no clear-cut
lattice symmetry present. In turn, we discard 'C'-type states.

\begin{figure}[tb]
	\includegraphics[width=\columnwidth]{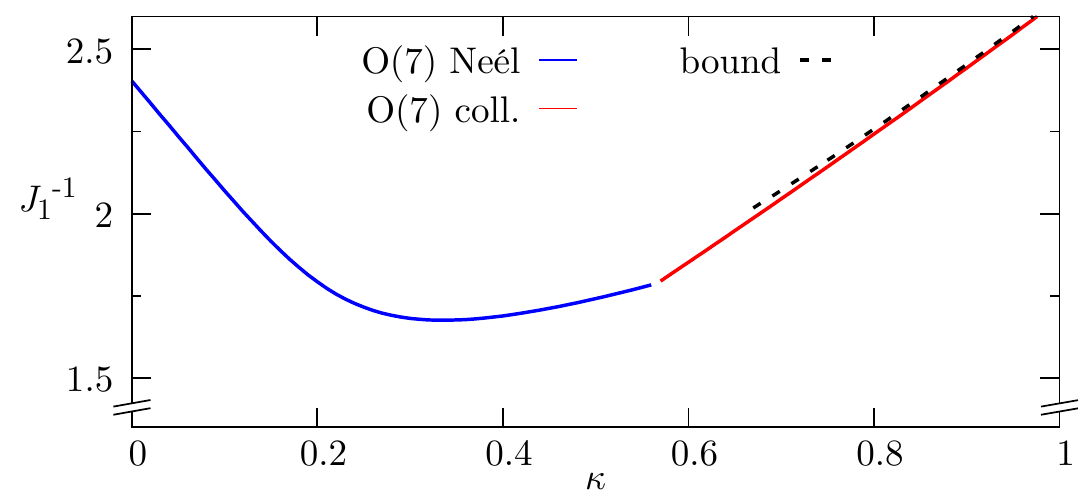}
	\caption{7th order results of phase diagram from Fig.~\ref{fig:4}a 	
		additionally with the point of gap closure in the two-particle sector 
		for series expansion at 7th and 5th order, for one- and two-particle 
		matrix elements, respectively.}
	\label{fig:7}
\end{figure}

Finally, in Fig.~\ref{fig:7} we discuss the options for bound-state criticality,
i.e., a gap-closure of the QDM phase comprising two-triplon bound states. To
appreciate this, we first realize, that for $\kappa\lesssim 0.5$ where the QDM
phase is expected to condense into N\'eel LRO, the ordering pitch vector is at
the $K$-point. There, and as can be seen from Fig.~\ref{fig:5}~(a), the
two-triplon bound states close to the $K$-point are at high energies. Comparing
the latter with those at the $K$-point in Fig.~\ref{fig:2}~(a) it is clear that
the breakdown of the QDM phase is driven by condensation of one-triplon states
only. This is very different for $\kappa\gtrsim 0.5$, where the pitch vector for
collinear LRO is at the $M$-point. Here, by comparing the gap-closures obtained
from the bare SE for the one-triplon states with the lowest two-triplon bound
states, i.e., 'B'-type, we are faced with the remarkable fact, that down to
$\kappa \sim 0.67$ one-triplon, as well as $S=0$ bound-states condense almost
simultaneously within a reasonable level of accuracy and with the bound-state
transition slightly above the magnetic one. Below $\kappa \sim 0.67$, a proper
separation of bound and continuum states from the $Q{=}2$-sector turns
infeasible. However it seems very likely that the black dashed line in
Fig.~\ref{fig:7} remains close to the red line, until the critical pitch
switches from collinear to N\'eel. Since 'B'-states have $S=0$, it is now very
tempting to speculate that these results indicate an additional, intervening
non-magnetic quantum phase between the QDM and collinear region, along its upper
edge in the quantum phase diagram of Fig. \ref{fig:4} (b). Even more surprising,
such a phase, separating QDM from magnetic spiral states has been found also in
the frustrated honeycomb bilayer Heisenberg model \cite{Zhang2018}, where the
intervening phase displayed nematic character. Consolidating our present
findings by resummation or higher-order SE remains an open question beyond this
work.

\section{Conclusion}
\label{sec:conclusion}

To conclude, using pCUT series expansion, we have investigated the elementary
excitations of the antiferromagnetic frustrated $J_1$-$J_2$ bilayer 
spin-$\sfrac{1}{2}$ Heisenberg model on the square lattice in its quantum dimer 
regime, with a particular focus on the stability of the dimer phase as well as on 
the spectrum of (anti)bound two-triplon states. We have determined the quantum 
critical lines for a condensation of the one-triplon excitation into magnetic 
phases. The location of these transitions is in good agreement with other published
analysis, starting from the magnetic phases. For a sizeable pocket in the
$J_1/J_2 - J_\perp$ plane, situated above the region of the potential quantum
spin-liquid of the single layer, we have provided an approximate upper bound for
its stability against the formation of a dimer gap. Regarding the two-triplon
spectrum we have uncovered a rich structure of collective (anti)bound states
which we have classified according to their total spin and rotational
symmetry. We found, that frustration impacts the continuum density of states
such, that not only the binding energy, but in particular the location of the
lowest lying bound states can be shifted strongly within the Brillouin zone. For
large frustration, this leads to a scenario in which the one-triplon and the
$S=0$ bound states condense very close to each other, with a slight preference
for the latter. This may suggests that the transition from the dimer into the
collinear magnetic state could involve an additional intermediate nonmagnetic
phase. This perspective calls for additional analysis. For an experimental probe
of our result in the two-triplon sector, future investigations of optical probes,
including phonon-assisted magnetic absorption, as well as magnetic Raman
scattering should be of interest.

\begin{acknowledgments}
This work has been supported in part by the DFG through Project A02 of SFB 1143
(Project-Id 247310070). Work of W.B. has been supported in part by the National
Science Foundation under Grant No. NSF PHY-1748958. W.B. also acknowledges kind
hospitality of the PSM, Dresden.
\end{acknowledgments}

\appendix

\section{Euler resummation}
\label{sec:appendix:euler}

Here we briefly describe the resummation of a power series using a variation of 
the Euler resummation.
An infinite power series in $x$ of the form
\begin{equation}
	f(x) = \sum_{n=0}^{\infty} a_n x^n
\end{equation}
can be rewritten as
\begin{equation}
	f(x) = \frac{1}{2}\sum_{n=0}^{\infty} 	\frac{1}{2^n} \sum_{k=0}^n 
	\binom{n}{k} a_k x^k\ ,
	\label{eq:euler}
\end{equation}
which is called the Euler resummation.
We are working with {\itshape finite} power series, depending on two parameters 
$J_1$ and $J_2$, which can be written as
\begin{align}
	E(J_1,J_2) &= \sum_{n=0}^{N_{\text{max}}} \sum_{l=0}^{n} \tilde{a}_{l,n-l} 
	J_1^l J_2^{n-l}
	\intertext{where $\tilde{a}_{l,n-l}$ are rational coefficients obtained by 
		pCUT and $N_{\text{max}}$ is the order of the SE. With the introduction 
		of 
		$\kappa = \frac{J_2}{J_1}$ this can be written as}
	E(J_1,\kappa) &= \sum_{n=0}^{N_{\text{max}}} a_{n}(\kappa) J_1^n
	\intertext{with}
	a_{n}(\kappa) &= \sum_{l=0}^{n} \tilde{a}_{l,n-l} \kappa^{n-l}\ .
\end{align}
Now $E(J_1,\kappa)$ is a polynomial in $J_1$ and the Euler resummation from
Eq.~\eqref{eq:euler} can be applied by 
limiting its $n$-sum to the same order $N_{\text{max}}$ as the bare series,
yielding a resummed series
\begin{equation}
	E_{\text{Euler}}(J_1,\kappa) = \frac{1}{2}\sum_{n=0}^{N_{\text{max}}} 	
	\frac{1}{2^n} \sum_{k=0}^n \binom{n}{k} a_k(\kappa) J_1^k\ .
\end{equation}
In practice this limits the impact of the highest order of the series 
expansion, counteracting an alternating behavior between orders (see 
Fig.~\ref{fig:3}~(b)). This can improve the convergence.


\begin{thebibliography}{99}

\bibitem{Sachdev2008}
S. Sachdev, 
Quantum Magnetism and Criticality, 
Nature Physics 4, 173 (2008).

\bibitem{Kitaev2006}
A. Kitaev, Ann. Phys. (N.Y.) \textbf{321}, 2 (2006).

\bibitem{Balents2010}
L. Balents,  Nature {\bf 464}, 199–208 (2010).

\bibitem{Henley2010}C. L. Henley, Annu. Rev. Condens. Matter Phys.
\textbf{1}, 179 (2010).

\bibitem{Castelnovo2012}C. Castelnovo, R. Moessner, and S. L. Sondhi,
Annual Review of Condensed Matter Physics \textbf{3}, 35 (2012). 

\bibitem{Misguich2012}
G. Misguich and C. Lhuillier,
in Frustrated Spin Systems (WORLD SCIENTIFIC, 2013,
Edited By: H T Diep), pp. 229–306.

\bibitem{Savary2016}L. Savary and L. Balents, Rep. Prog. Phys. \textbf{80},
016502 (2017).

\bibitem{Sachdev2018}S. Sachdev, Rep. Prog. Phys. \textbf{82}, 014001
(2018).

\bibitem{Chandra1988}
P. Chandra and B. Dou\c{c}ot,
Phys. Rev. B {\bf 38}, 9335 (1988).


\bibitem{Henley1989}
C. L. Henley,
Phys. Rev. Lett. {\bf 62}, 2056 (1989).

\bibitem{Moreo1990}
A. Moreo, E. Dagotto, T. Jolicoeur and J. Riera,
Phys. Rev. B {\bf 42}, 6283 (1990).

\bibitem{Chandra1990}
P. Chandra, P. Coleman, and A. Larkin,
Phys. Rev. Lett. {\bf 64}, 88 (1990).

\bibitem{Luttinger1946}
J. M. Luttinger and L. Tisza,
Phys. Rev. {\bf 70}, 954 (1946).


\bibitem{Elbio1989}
E. Dagotto and A. Moreo,
Phys. Rev. Lett. {\bf 63}, 2148 (1989).

\bibitem{Didier1991}
D. Poilblanc, E. Gagliano, S. Bacci, and E. Dagotto
Phys. Rev. B {\bf 43}, 10970 (1991).

\bibitem{Didier1996}
H. J. Schulz, T. A. L. Ziman, and D. Poilblanc,
J. Phys. I {\bf 6}, 675 (1996).

\bibitem{Didier2006}
M. Mambrini, A. Läuchli, D. Poilblanc, and F. Mila,
Phys. Rev. B {\bf 74}, 144422 (2006).


\bibitem{Oitmaa1996}
J. Oitmaa and Zheng Weihong
Phys. Rev. B {\bf 54}, 3022 (1996).

\bibitem{Rajiv1999}
R. R. P. Singh, Z. Weihong, C. J. Hamer, and J. Oitmaa,
Phys. Rev. B {\bf 60}, 7278 (1999).

\bibitem{Sirker2006}
J. Sirker, Z. Weihong, O. P. Sushkov, and J. Oitmaa,
Phys. Rev. B {\bf 73}, 184420 (2006).


\bibitem{Darradi2008}
R. Darradi, O. Derzhko, R. Zinke, J. Schulenburg, S. E. Krüger, and J. Richter,
Phys. Rev. B {\bf 78}, 214415 (2008).

\bibitem{Richter2015}
J. Richter, R. Zinke, and D. J. J. Farnell,
Eur. Phys. J. B {\bf 88}, 2 (2015).

\bibitem{Papastathopoulos2022}
A. Papastathopoulos-Katsaros, C. A. Jim\'enez-Hoyos, T. M. Henderson,
and G. E. Scuseria,
J. Chem. Theory Comput. 2022, {\bf 18}, 4293 (2022).


\bibitem{Mezzacapo2012}
F. Mezzacapo,
Phys. Rev. B {\bf 86}, 045115 (2012).

\bibitem{Ren2014}
Y.-Z. Ren, N.-H. Tong, and X.-C. Xie,
J. Phys.: Cond. Matt. {\bf 26}, 115601 (2014).


\bibitem{Jiang2012}
H.-C. Jiang, H. Yao, and L. Balents,
Phys. Rev. B {\bf 86}, 024424 (2012).

\bibitem{Gong2014}
S.-S. Gong, W. Zhu, D. N. Sheng, O. I. Motrunich, and M. P. A. Fisher,
Phys. Rev. Lett. {\bf 113}, 027201 (2014).

\bibitem{Wang2018}
L. Wang and A. W. Sandvik,
Phys. Rev. Lett. {\bf 121}, 107202 (2018).


\bibitem{Haghshenas2018}
R. Haghshenas and D. N. Sheng,
Phys. Rev. B {\bf 97}, 174408 (2018).

\bibitem{Hasik2021}
J. Hasik, D. Poilblanc, and F. Becca,
SciPost Phys. {\bf 10}, 012 (2021).

\bibitem{Liu2022}
W.-Y. Liu, S.-S. Gong, Y.-B. Li, D. Poilblanc, W.-Q. Chen,
and Z.-C. Gu, Science Bulletin {\bf 67}, 1034 (2022).


\bibitem{Hering2019}
M. Hering, J. Sonnenschein, Y. Iqbal and J. Reuther,
Phys. Rev. B {\bf 99}, 100405(R) (2019).

\bibitem{Roscher2019}
D. Roscher, N. Gneist, M. M. Scherer, S. Trebst, and S. Diehl,
Phys. Rev. B {\bf 100}, 125130 (2019).


\bibitem{Zhitomirsky1996}
M. E. Zhitomirsky and K. Ueda,
Phys. Rev. B {\bf 54}, 9007 (1996).

\bibitem{Doretto2014}
R. L. Doretto,
Phys. Rev. B {\bf 89}, 104415 (2014).


\bibitem{Hu2013}
W.-J. Hu, F. Becca, A. Parola and S. Sorella,
Phys. Rev. B {\bf 88}, 060402(R) (2013).

\bibitem{Morita2015}
S. Morita, R. Kaneko, and M. Imada,
J. Phys. Soc. Jap. {\bf 84}, 024720 (2015).

\bibitem{Ferrari2020}
F. Ferrari and F. Becca,
Phys. Rev. B {\bf 102}, 014417 (2020).

\bibitem{Nomura2021}
Y. Nomura and M. Imada,
Phys. Rev. X {\bf 11}, 031034 (2021).


\bibitem{Todate2007}
Y. Todate, W. Higemoto, K. Nishiyama and K. Hirota,
J. Phys. Chem. Solids {\bf 68}, 2107 (2007).

\bibitem{Vasala2014a}
S. Vasala, H. Saadaoui, E. Morenzoni, O. Chmaissem, T.-S. Chan, J.-M. Chen,
Y.-Y. Hsu, H. Yamauchi and M. Karppinen,
Phys. Rev. B {\bf 89}, 134419 (2014).

\bibitem{Vasala2014b}
S. Vasala, M. Avdeev, S. Danilkin, O. Chmaissem and M. Karppinen,
Magnetic structure of Sr$_2$CuWO$_6$,
J. Phys.: Condens. Matter {\bf 26}, 496001 (2014).

\bibitem{Koga2016}
T. Koga, N. Kurita, M. Avdeev, S. Danilkin, T. J. Sato and H. Tanaka,
Phys. Rev. B {\bf 93}, 054426 (2016).

\bibitem{Melzi2000}
R. Melzi, P. Caretta, A. Lascialfari, M. Mambrini, M. Troyer, P. Millet and F. 
Mila,
Phys. Rev. Lett. {\bf 85}, 1318 (2000).

\bibitem{Melzi2001}
R. Melzi, S. Aldrovandi, F. Tedoldi, P. Carretta, P. Millet and F. Mila,
Phys. Rev. B {\bf 64}, 024409 (2001).


\bibitem{Elbio1996}
E. Dagotto and T. M. Rice,
Science {\bf 271}, 618 (1996).

\bibitem{Wang2006}
L. Wang, K. S. D. Beach, and A. W. Sandvik,
Phys. Rev. B {\bf 73}, 014431 (2006).

\bibitem{Matsumoto2004}
M. Matsumoto, B. Normand, T. M. Rice, and M. Sigrist,
Phys. Rev. B {\bf 69}, 054423 (2004).


\bibitem{Zhang2016}
H. Zhang, C. A. Lamas, M. Arlego, and W. Brenig
Phys. Rev. B {\bf 93}, 235150 (2016).

\bibitem{Zhang2018}
H. Zhang, C.A. Lamas, M. Arlego and W. Brenig,
Phys. Rev. B {\bf 97}, 235123 (2018).

\bibitem{Seifert2018}
U. F. P. Seifert, J. Gritsch, E. Wagner, D. G. Joshi, W. Brenig,
M. Vojta, and K. P. Schmidt,
Phys. Rev. B {\bf 98}, 155101 (2018).

\bibitem{Joshi2019}
D. G. Joshi and A. P. Schnyder,
Phys. Rev. B {\bf 100}, 020407(R) (2019).

\bibitem{Acevedo2021}
S. Acevedo, C. A. Lamas, and P. Pujol,
Phys. Rev. B {\bf 104}, 214412 (2021).

\bibitem{Szalowski2012}
K. Szalowski and T. Balcerzak,
Physica A (Amsterdam) {\bf 391}, 2197 (2012).

\bibitem{Hoermann2020}
M. Hörmann and K. P. Schmidt,
Phys. Rev. B {\bf 102}, 094427 (2020).

\bibitem{Nyhegn2022}
J. H. Nyhegn, K. K. Nielsen, and G. M. Bruunm
Phys. Rev. B {\bf 106}, 155160 (2022).

\bibitem{Wagner2021}
E. Wagner and W. Brenig,
Phys. Rev. B {\bf 104}, 115123 (2021).

\bibitem{Ghader2021}
D. Ghader,
New J. Phys. {\bf 23}, 053022 (2021).


\bibitem{Sebastian2006}
S. E. Sebastian, N. Harrison, C. D. Batista, L. Balicas, M. Jaime, P. A. Sharma,
N. Kawashima, and I. R. Fisher,
Nature {\bf 441}, 617 (2006).

\bibitem{Allenspach2021}
S. Allenspach, P. Puphal, J. Link, I. Heinmaa, E. Pomjakushina, C. Krellner,
J. Lass, G. S. Tucker, C. Niedermayer, S. Imajo, Y. Kohama, K. Kindo,
S. Krämer, M. Horvatić, M. Jaime, A. Madsen, A. Mira, N. Laflorencie, F. Mila,
B. Normand, C. Rüegg, R. Stern, and F. Weickert,
Phys. Rev. Res. {\bf 3}, 023177 (2021).

\bibitem{Merchant2014}
P. Merchant, B. Normand, K. W. Krämer, M. Boehm, D. F. McMorrow, and Ch. Rüegg,
Nature Physics {\bf 10}, 373 (2014).

\bibitem{Stone2008}
M. B. Stone, M. D. Lumsden, S. Chang, E. C. Samulon, C. D. Batista, and I. R. 
Fisher,
Phys. Rev. Lett. {\bf 100}, 237201 (2008), ibid {\bf 105}, 169901 (2010).

\bibitem{Kageyama1999}
H. Kageyama, K. Yoshimura, R. Stern, N. V. Mushnikov, K. Onizuka, M. Kato,
K. Kosuge, C. P. Slichter, T. Goto, and Y. Ueda,
Phys. Rev. Lett. {\bf 82}, 3168 (1999).


\bibitem{Hida1996}
K. Hida,
J. Phys. Soc. Jpn. {\bf 65}, 594 (1996).

\bibitem{Hida1998}
K. Hida,
J. Phys. Soc. Jpn {\bf 67}, 1540 (1998).

\bibitem{Bishop2019}
R. F. Bishop, P. H. Y. Li, O. G\"otze and J. Richter,
Phys. Rev. B {\bf 100}, 024401 (2019).

\bibitem{Knetter2000a}
C. Knetter, G.S. Uhrig,
Eur. Phys. J. B {\bf 13}, 209 (2000).

\bibitem{Wegner1994}
F. Wegner,
Ann. Physik {\bf 3}, 77 (1994).

\bibitem{Windt2001}
M. Windt, M. Grüninger, T. Nunner, C. Knetter, K. P. Schmidt, G. S. Uhrig,
T. Kopp, A. Freimuth, U. Ammerahl, B. Büchner, and A. Revcolevschi,
Phys. Rev. Lett. {\bf 87}, 127002 (2001).

\bibitem{Arlego2013}
M. Arlego, W. Brenig, Y. Rahnavard, B. Willenberg,
H. D. Rosales, and G. Rossini,
Phys. Rev. B {\bf 87}, 014412 (2013).

\bibitem{Brenig2002}
W. Brenig and A. Honecker
Phys. Rev. B {\bf 65}, 140407(R) (2002).



\bibitem{Knetter2003t}
C. Knetter, 
{\itshape Perturbative continuous unitary transformations: spectral 	
properties of low dimensional spin systems}, 
Doctoral dissertation, University of Cologne, 2003.

\bibitem{Zheng1997}
W. Zheng,
Phys. Rev. B {\bf 55}, 12267 (1997).

\bibitem{Knetter2003}
C. Knetter, K.P. Schmidt, G.S. Uhrig,
Eur. Phys. J. B {\bf 36}, 525-544 (2003).


\bibitem{Collins2008}
A. Collins and C. J. Hamer, 
Phys. Rev. B {\bf 78}, 054419 (2008).

\end{thebibliography}
\end{document}